\documentclass[fleqn]{llncs}

\usepackage{amsmath,amssymb}
\usepackage{hyperref}

\hypersetup{%
  pdftitle={Characterizations and Kullback-Leibler Divergence of Gompertz Distributions},
  pdfauthor={Christian Bauckhage},
  pdfsubject={data science, statistics},
  pdfkeywords={Gompertz distribution, Kullback-Leibler divergence, Gumbel distribution, Frechet distribution, Weibull distribution},
  colorlinks=true,
  urlcolor=blue,
  citecolor=red,
  bookmarks=false}

\title{Characterizations and Kullback-Leibler Divergence of Gompertz Distributions}

\author{Christian Bauckhage}

\institute{%
  B-IT, University of Bonn, Bonn, Germany \\
  Fraunhofer IAIS, Sankt Augustin, Germany \\
  \email{http://mmprec.iais.fraunhofer.de/bauckhage.html}}

\begin{document}

\maketitle

\begin{abstract}
In this note, we characterize the Gompertz distribution in terms of extreme value distributions and point out that it implicitly models the interplay of two antagonistic growth processes. In addition, we derive a closed form expressions for the Kullback-Leibler divergence between two Gompertz Distributions. Although the latter is rather easy to obtain, it seems not to have been widely reported before.
\end{abstract}

\section{The Gompertz Distribution}

The \href{http://en.wikipedia.org/wiki/Gompertz_distribution}{Gompertz distribution} provides a statistical formulation of the Gompertz law of mortality \cite{Winsor1932-TGC}. Its probability density function (pdf) is defined for $x \in [0,\infty)$ and given by
\begin{equation}
  \label{eq:GompertzPDF}
  f(x \mid b, q) = e^q \, b \, q \, e^{b x} \, e^{-q \, e^{bx}}
\end{equation}
where the parameter $b > 0$ determines scale and $q > 0$ is a shape parameter. The corresponding cumulative density function (cdf) amounts to
\begin{equation}
\label{eq:GompertzCDF}
  F(x \mid b, q) = 1 - e^q \, e^{-q \, e^{bx}}
\end{equation}
and will be of interest in our discussion below. 

Regarding the density in \eqref{eq:GompertzPDF}, we note that it is unimodal and rather flexible. Depending on the choice of $b$ and $q$, it may be skewed to the left or to the right; however, for $q \geq 1$, its mode will always be at $0$.

Due to its origins as a model of mortality, the Gompertz distribution is a staple in statistical biology and the demographic and actuarial sciences \cite{Strehler1960-GTO,Banks1994-GAD}. It was observed to model income distributions \cite{Mouyra2009-EFT} and has been used as a model of the diffusion of novel products as well as of customer life-time values \cite{Bemmaor1994-MTD,Jaakola1996-CAA,Geroski2000-MOT} in economics and marketing . Finally, in the context of social media analysis, the Gompertz distribution was found to account well for the temporal evolution of collective attention to viral Web content or social media services \cite{Bauckhage2013-MMO,Bauckhage2014-CAT}.

Our goal with this note is to provide an accessible account of some of the properties of the Gompertz distribution. Furthermore, we derive a closed form expression for the Kullback-Leibler divergence between Gompertz distributions which is useful for the purpose of model selection or statistical inference.

\section{Interpretation in Terms of Extreme Value Distributions}

Interestingly, the Gompertz distribution is rather closely related to extreme value theory. Here, we briefly demonstrate that it can be expressed in terms of the three extreme value distributions.

\subsubsection{First of all,} the Gompertz distribution corresponds to a zero-truncated Gumbel minimum distribution.

The \href{http://en.wikipedia.org/wiki/Gumbel_distribution}{Gumbel distribution} is the type I extreme value distribution. When used to model the distribution of sample minima, its pdf is defined for $x \in (-\infty, \infty)$ and usually expressed as
\begin{equation}
\label{eq:Gumbel}
f_{\mathcal{G}}(x \mid m, s)
= \frac{1}{s} \, e^{(x-m)/s} \, e^{- e^{(x-m)/s}}
= \frac{1}{s} \, e^{\frac{x}{s}} \, e^{-\frac{m}{s}} \, e^{- e^{\frac{x}{s}} \, e^{-\frac{m}{s}}}
\end{equation}
where $m$ is a location parameter and $s > 0$ determines scale. Hence, defining $b = \tfrac{1}{s}$ and $q = e^{-m/s}$ allows us to re-parameterize \eqref{eq:Gumbel} and to write it as
\begin{equation}
\label{eq:Gumbel2PDF}
f_{\mathcal{G}}(x \mid b, q) = b \, q \, e^{bx} \, e^{-q \, e^{bx}}
\end{equation}
such that the corresponding cumulative density function amounts to
\begin{equation}
\label{eq:Gumbel2CDF}
F_{\mathcal{G}}(x \mid b, q) = 1 - e^{-q \, e^{bx}}.
\end{equation}
Looking at the cumulative density in \eqref{eq:Gumbel2CDF}, we note that $\lim_{x \rightarrow \infty} F_{\mathcal{G}}(x) = 1$ as well as $F_{\mathcal{G}}(0) = 1 - e^{-q}$. Accordingly, by left truncating the density in \eqref{eq:Gumbel2PDF} at $0$, we obtain a distribution whose pdf is given by
\begin{equation}
\frac{f_{\mathcal{G}}(x)}{\int_0^\infty f_{\mathcal{G}}(x) dx} = \frac{f_{\mathcal{G}}(x)}{1 - F_{\mathcal{G}}(0)}
= \frac{f_{\mathcal{G}}(x)}{1 - (1 - e^{-q})} = e^{q} f_{\mathcal{G}}(x) = e^q \, b \, q \, e^{bx} \, e^{-q \, e^{bx}}.
\end{equation}
This, however, is indeed the probability density of the Gompertz distribution as introduced in \eqref{eq:GompertzPDF}.

\subsubsection{Second of all,} the Gompertz is indirectly related to the Fr\'echet and to the Weibull distribution.

The \href{http://en.wikipedia.org/wiki/Fr%C3%A9chet_distribution}{Fr\'echet distribution} is the type II extreme value distribution. It is usually defined for $x \in (0, \infty)$ in which case its pdf is given by
\begin{equation}
\label{eq:FrechetPDF}
f_{\mathcal{F}}(x \mid a, r) = \frac{a}{r} \, \left( \frac{x}{r}\right)^{-1-a} \, e^{-\left( \frac{x}{r} \right)^{-a}}
\end{equation}
where $a > 0$ and $r > 0$ are shape and scale parameters, respectively.

The \href{http://en.wikipedia.org/wiki/Weibull_distribution}{Weibull distribution} is the type III extreme value distribution. It is commonly defined for $x \in [0, \infty)$ and its pdf amount to
\begin{equation}
\label{eq:WeibullPDF}
f_{\mathcal{W}}(x \mid k, l) = \frac{k}{l} \, \left( \frac{x}{l}\right)^{k-1} \, e^{-\left( \frac{x}{l} \right)^{k}}
\end{equation}
where $k > 0$ and $l > 0$ are shape and scale parameters, respectively.

In order to expose the connections between the densities in \eqref{eq:FrechetPDF} and \eqref{eq:WeibullPDF} and the Gompertz density in \eqref{eq:GompertzPDF}, we recall that if a random variable $X$ is distributed according to $f_X(x)$, the monotonously transformed random variable $Y = h(X)$ has a pdf that is given by
\begin{equation}
f_Y(y) = f_X \bigl( h^{-1}(y) \bigr) \left \lvert \frac{d}{dy} h^{-1} (y) \right \rvert.
\end{equation}
Using this identity, it is straightforward to see that the Gompertz distribution also results from transforming Fr\'echet or Weibull distributions.

In particular, if $f_X(x)$ is a Fr\'echet density and $y = - \ln x$, then
\begin{equation}
x = e^{-y} \quad \text{and} \quad \frac{dx}{dy} = - e^{-y}
\end{equation}
so that
\begin{align}
f_Y(y)
& = \frac{a}{r} \, \left( \frac{e^{-y}}{r}\right)^{-1-a} \, e^{-\left( \frac{e^{-y}}{r} \right)^{-a}} \, \left\lvert -e^{-y} \right \rvert \\
& = \frac{a}{r} \, r^{a+1} \, e^{-y(-a-1) -y} \, e^{-r^a \, e^{ya}} \\
& = a \, r^a \, e^{ya} \, e^{-r^a \, e^{ya}} \label{eq:Frechet-subst} \\
& = b \, q \, e^{b y} \, e^{-q \, e^{b y}} \label{eq:logFrechet}
\end{align}
where, in \eqref{eq:Frechet-subst}, we substituted $b = a$ and $q = r^a $. The expression in \eqref{eq:logFrechet}, however, is the Gumbel density known from  \eqref{eq:Gumbel2PDF}. This is to say that the Gompertz density $f(y) = e^q \, f_Y(y)$ is a truncated, negative log-transformed Fr\'echet density.

By the same token, if $f_X(x)$ is a Weibull density and $y = \ln x$, then
\begin{equation}
x = e^y \quad \text{and} \quad \frac{dx}{dy} = e^y
\end{equation}
so that
\begin{align}
f_Y(y)
& = \frac{k}{l} \, \left( \frac{e^{y}}{l}\right)^{k-1} \, e^{-\left( \frac{e^{y}}{l} \right)^{k}} \, \left\lvert e^{y} \right \rvert \\
& = \frac{k}{l^k} \, e^{yk} \, e^{- \frac{1}{l^k} e^{yk}} \label{eq:Weibull-subst} \\
& = b \, q \, e^{b y} \, e^{-q \, e^{b y}} \label{eq:logWeibull}
\end{align}
where, this time, we have substituted $b = k$ and $q = \tfrac{1}{l^k}$ in \eqref{eq:Weibull-subst}. The expression in \eqref{eq:logWeibull} corresponds once more to the Gumbel pdf in \eqref{eq:Gumbel2PDF}. This is to say that the Gompertz density $f(y) = e^q \, f_Y(y)$ is also a truncated log-transformed Weibull density.



\section{Interpretation in Terms of Antagonistic Growth Dynamics}

If we consider the cumulative density function of the Gompertz distribution as introduced in \eqref{eq:GompertzPDF}, we note that
\begin{equation}
  e^q \, e^{-q \, e^{bx}} = 1 -  F(x \mid b, q).
\end{equation}
This expression allows for yet another, physically meaningful interpretation of the Gompertz distribution. In particular, plugging this expression into \eqref{eq:GompertzPDF} yields
\begin{align}
f(x \mid b, q)
& = b \, q \, e^{b x} \, \bigl( 1 -  F(x \mid b, q) \bigr) \notag \\
& = b \, q \, e^{b x} - b \, q \, e^{b x} \,  F(x \mid b, q)
\end{align}
and we recognize that the Gompertz pdf implicitly models a subtractive growth dynamic. In other words, the Gompertz distribution can be understood as a growth model that combines a term
\begin{equation}
f_g(x) = b \, q \, e^{b x}
\end{equation}
with a term
\begin{equation}
f_d(x) = b \, q \, e^{b x} \,  F(x \mid b, q).
\end{equation}
If the variable $x$ is understood to represent time, $f_g$ can be interpreted as the propensity of an entity to \emph{grow} while $f_d$ can be seen as the propensity for the entity to \emph{decline}. Both, growth and decline, depend exponentially on $x$ but counteract each other. If, for instance, the entity in question is the amount of attention paid to a novelty, the component $F(x \mid b, q)$ of $f_d$ can be understood as the relative amount of attention the novelty has received so far. As $F$ grows monotonously from $0$ to $1$, we see that, for small values of $x$, the propensity $f_g$ for growth will exceed the propensity $f_d$ for decline. For growing $x$, however, the propensity for decline will approach the propensity for growth. As both dynamics are coupled in a subtractive manner, this means that the overall dynamic $f(x)$ will be characterized by an initial phase of rising attention to the novelty followed by a prolonged phase of demise.

Interestingly, since the factor $F(x \mid b, q)$ in $f_d$ can be understood as the amount of attention received up until $x$, the speed of decline in attention apparently depends on the overall novelty of whatever attention is paid to. Temporal dynamics like these are known to characterize the evolution of \emph{fads} \cite{Meyerson1957-NOA}. Seen from this point of view, it is thus noteworthy and revealing that the Gompertz distribution has been found to accurately model general trends in time series that indicate collective interest in topics and services on the Web \cite{Bauckhage2013-MMO,Bauckhage2014-CAT}.

\section{Computing the Kullback-Leibler Divergence between Gompertz Distributions}

The Kullback-Leibler (KL) divergence or relative entropy provides measures the similarity of two probability distributions $P$ and $Q$
\cite{Kullback1951-OIA}. In case both distributions are continuous, it
is defined as
\begin{equation}
  \label{eq:DKL}
  D_{KL} \bigl( P \parallel Q \bigr) = \int\limits_{-\infty}^{\infty} p(x) \, \ln \frac{p(x)}{q(x)} \, dx
\end{equation}
where $p(x)$ and $q(x)$ denote the corresponding probability density functions. The KL divergence can be understood as the information loss if $P$ is modeled in terms of $Q$. Accordingly, the smaller
$D_{KL} \bigl( P \parallel Q \bigr)$, the more similar are $P$ and $Q$. Although this is akin to the behavior of a distance, we note that the KL divergence should not be confused with a distance since it is neither symmetric nor satisfies the triangle inequality.

\subsection{Step by Step Solution}

Plugging two Gompertz distributions $F_1$ and $F_2$ into \eqref{eq:DKL} and noting once again that their densities are defined for $x \in [0,
\infty)$ yields
\begin{equation}
  \label{eq:DKLGO}
  D_{KL} \bigl( F_1 \parallel F_2 \bigr) = \int\limits_{0}^{\infty} f_1(x \mid b_1, q_1) \, \ln \frac{f_1(x \mid b_1, q_1)}{f_2(x \mid b_2, q_2)} dx.
\end{equation}

We begin evaluating this expression by considering the logarithmic factor inside the integral. Given the definition of the Gompertz distribution in \eqref{eq:GompertzPDF}, we may write it as 
\begin{equation}
\label{eq:DKLlog}
\ln \left[ \underbrace{\frac{e^{q_1} \, b_1 \, q_1}{e^{q_2} \, b_2 \, q_2}}_{A} \cdot \frac{e^{b_1 x}}{e^{b_2 x}} \cdot \frac{e^{-q_1 e^{b_1 x}}} {e^{-q_2 e^{b_2 x}}} \right]
= \ln A + x \, (b_1 - b_2) + q_2 \, e^{b_2 x} - q_1 \, e^{b_1 x}
\end{equation}
and observe that the term $\ln A$ is a constant independent of the variable of integration $x$. Plugging \eqref{eq:DKLlog} back into \eqref{eq:DKLGO}, we therefore obtain
\begin{align}
  & \int\limits_{0}^{\infty} f_1(x \mid b_1, q_1) \, \ln A + x \, (b_1 - b_2) + q_2 \, e^{b_2 x} - q_1 \, e^{b_1 x} \, dx \notag \\
= & \int\limits_{0}^{\infty} f_1(x \mid b_1, q_1) \, \ln A \, dx \label{eq:intconst} \\
  & + \int\limits_{0}^{\infty} f_1(x \mid b_1, q_1) \, x \, (b_1 - b_2) \, dx \label{eq:intb1b2} \\
  & + \int\limits_{0}^{\infty} f_1(x \mid b_1, q_1) \, q_2 \, e^{b_2 x} \, dx \label{eq:intq2} \\
  & - \int\limits_{0}^{\infty} f_1(x \mid b_1, q_1) \, q_1 \, e^{b_1 x} \, dx. \label{eq:intq1}
\end{align}

Next, we evaluate the integrals in \eqref{eq:intconst} to \eqref{eq:intq2} one by one and then assemble the final solution from the intermediate results we thus obtain.

\subsubsection{Solving \eqref{eq:intconst}} Since $f_1(x \mid b_1, q_1)$ is a probability density over $[0,\infty)$, it is obvious that
\begin{equation}
\int\limits_{0}^{\infty} f_1(x \mid b_1, q_1) \, \ln A \, dx  = 1 \cdot \ln A = \ln A.
\end{equation}
In other words, the term in \eqref{eq:intconst} evaluates to
\begin{equation}
\label{eq:intconst-final}
\ln \frac{e^{q_1} \, b_1 \, q_1}{e^{q_2} \, b_2 \, q_2}.
\end{equation}

\subsubsection{Solving \eqref{eq:intb1b2}} Plugging the definition for $f_1(x \mid b_1, q_1)$ into \eqref{eq:intb1b2} yields
\begin{align}
& \int\limits_{0}^{\infty} e^{q_1} \, b_1 \, q_1 \, e^{b_1 x} \, e^{-q_1 \, e^{b_1 x}} \, x \, (b_1 - b_2) \, dx \notag \\
= & \; e^{q_1} \, q_1 \, (b_1 - b_2) \, \int\limits_{0}^{\infty} b_1 \, e^{b_1 x} \, e^{-q_1 \, e^{b_1 x}} x \; dx. \label{eq:intb1b2-intermediate}
\end{align}
In order to solve the integral on the right hand side of \eqref{eq:intb1b2-intermediate}, we consider the following substitution
\begin{equation}
\label{eq:subst}
y = e^{b_1 x}.
\end{equation}
Accordingly, we have that
\begin{equation}
dy = b_1 \, e^{b_1 x} \, dx
\end{equation}
as well as
\begin{equation}
x = \frac{1}{b_1} \, \ln y.
\end{equation}
Using these identities then leads to
\begin{equation}
\label{eq:intb1b2-intermediate2}
e^{q_1} \, q_1 \, (b_1 - b_2) \, \int\limits_{0}^{\infty} b_1 \, e^{b x} \, e^{-q_1 \, e^{b_1 x}} x \; dx
= \frac{e^{q_1} \, q_1 \, (b_1 - b_2)}{b_1} \, \int\limits_{1}^{\infty} \ln y \, e^{-q_1 y} \, dy
\end{equation}
and we note that the substitution causes a change of the limits of integration. Regarding the integral on the right hand side, we consult \cite[eq. 4.331]{Gradshteyn2007-TOI} and find
\begin{equation}
\label{eq:intb1b2-intermediate3}
\int\limits_{1}^{\infty} \ln y \, e^{-q_1 y} \, dy = - \frac{1}{q_1} \operatorname{Ei}(- q_1)
\end{equation}
where $\operatorname{Ei} (\cdot)$ denotes the \href{http://en.wikipedia.org/wiki/Exponential_integral}{exponential integral} and we recall that this \href{http://en.wikipedia.org/wiki/Special_function}{special function} is defined as
\begin{equation}
\operatorname{Ei}(x) = - \int\limits_{-x}^{\infty} \frac{e^{-t}}{t} \, dt = \int\limits_{-\infty}^{x} \frac{e^{t}}{t} \, dt.
\end{equation}
where $x < 0$.

Hence, multiplying the result in \eqref{eq:intb1b2-intermediate3} with the factor on the right hand side of \eqref{eq:intb1b2-intermediate2} provides our next intermediate result, namely, that the \eqref{eq:intb1b2} evaluates to
\begin{equation}
\label{eq:intb1b2-final}
e^{q_1} \left(\frac{b_2}{b_1} - 1 \right) \, \operatorname{Ei}(- q_1).
\end{equation}

\subsubsection{Solving \eqref{eq:intq2}} Plugging the definition for $f_1(x \mid b_1, q_1)$ into \eqref{eq:intq2} yields
\begin{align}
\int\limits_{0}^{\infty} e^{q_1} \, b_1 \, q_1 \, e^{b_1 x} \, e^{-q_1 \, e^{b_1 x}} \, q_2 \, e^{b_2 x} \, dx
= & e^{q_1} \, q_1 \, q_2 \, \int\limits_{0}^{\infty} b_1 \, e^{b_1 x} \, e^{-q_1 \, e^{b_1 x}} \, e^{b_2 x} \, dx \\
= & e^{q_1} \, q_1 \, q_2 \, \int\limits_{1}^{\infty} y^{\frac{b_2}{b_1}} \, e^{-q_1 y} \, dy \label{eq:intq2-intermediate}
\end{align}
where \eqref{eq:intq2-intermediate} results from applying the substitution which we introduced in \eqref{eq:subst}. Consulting \cite[eq. 4.381]{Gradshteyn2007-TOI}, we find that the integral in \eqref{eq:intq2-intermediate} evaluates to
\begin{equation}
\label{eq:intq2-intermediate2}
\int\limits_{1}^{\infty} y^{\frac{b_2}{b_1}} \, e^{-q_1 y} \, dy = q_1^{-\left(\frac{b_2}{b_1}+1\right)} \, \Gamma \left(\frac{b_2}{b_1}+1, q_1 \right)
\end{equation}
where $\Gamma(\cdot, \cdot)$ denotes the upper \href{http://en.wikipedia.org/wiki/Incomplete_gamma_function}{incomplete gamma function} for which we recall that it is defined as
\begin{equation}
\Gamma(s, x) = \int\limits_{x}^{\infty} t^{s-1} e^{-t} \, dt.
\end{equation}

Accordingly, multiplying the result in \eqref{eq:intq2-intermediate2} with the factor on the right hand side of \eqref{eq:intq2-intermediate} yields the next intermediate result; the term in \eqref{eq:intq2} amounts to
\begin{equation}
\label{eq:intq2-final}
e^{q_1} \, q_2 \, q_1^{-\frac{b_2}{b_1}} \, \Gamma \left(\frac{b_2}{b_1}+1, q_1 \right).
\end{equation}

\subsubsection{Solving \eqref{eq:intq1}} Finally, plugging a Gompertz density $f_1(x \mid b_1, q_1)$ into \eqref{eq:intq1} yields
\begin{equation}
\label{eq:intq1-intermediate}
- \int\limits_{0}^{\infty} e^{q_1} \, b_1 \, q_1 \, e^{b_1 x} \, e^{-q_1 \, e^{b_1 x}} \, q_1 \, e^{b_1 x} \, dx
= - e^{q_1} \, q_1^2 \, \int\limits_{1}^{\infty} y \, e^{-q_1 y} \, dy
\end{equation}
where we have once again made use of the substitution in \eqref{eq:subst}. For the integral on the right hand side of \eqref{eq:intq1-intermediate}, we have
\begin{equation}
\int\limits_{1}^{\infty} y \, e^{-q_1 y} \, dy
= \left[ - \frac{e^{-q_1 x}}{q_1^2} \, (q_1 \, x + 1)\right]_1^\infty
= \frac{e^{-q_1}}{q_1^2} \, (q_1 + 1)
\end{equation}
so that the term in \eqref{eq:intq1} simplifies to
\begin{equation}
\label{eq:intq1-final}
- (q_1 + 1).
\end{equation}

\subsection{Final Result}

Assembling the four intermediate results in \eqref{eq:intconst-final}, \eqref{eq:intb1b2-final}, \eqref{eq:intq2-final}, and \eqref{eq:intq1-final} establishes that: The KL divergence between two Gompertz densities $f_1$ and $f_2$ amounts to
\begin{align}
  & \int\limits_{0}^{\infty} f_1(x \mid b_1, q_1) \, \ln \frac{f_1(x \mid b_1, q_1)}{f_2(x \mid b_2, q_2)} dx \\
= & \ln \frac{e^{q_1} \, b_1 \, q_1}{e^{q_2} \, b_2 \, q_2}
+ e^{q_1} \left[ \left(\frac{b_2}{b_1} - 1 \right) \, \operatorname{Ei}(- q_1)
                 + \frac{q_2}{q_1^{\frac{b_2}{b_1}}} \, \Gamma \left(\frac{b_2}{b_1}+1, q_1 \right) \right]
- (q_1 + 1).
\end{align}

\bibliographystyle{splncs}
\bibliography{literature}

\begin{thebibliography}{10}

\bibitem{Winsor1932-TGC}
Winsor, C.:
\newblock {The Gompertz Curve as a Growth Curve}.
\newblock PNAS \textbf{18}(1) (1932)  1--8

\bibitem{Strehler1960-GTO}
Strehler, B., Mildvan, A.:
\newblock {General Theory of Mortality and Aging}.
\newblock Science \textbf{132}(3418) (1960)  14--21

\bibitem{Banks1994-GAD}
Banks, R.:
\newblock {Growth and Diffusion Phenomena}.
\newblock Springer (1994)

\bibitem{Mouyra2009-EFT}
Moura, N., Ribeiro, M.:
\newblock {Evidence for the Gompertz Curve in the Income Distribution of Brazil
  1978–2005}.
\newblock European Physical Journal B \textbf{67}(1) (2009)  101--120

\bibitem{Bemmaor1994-MTD}
Bemmaor, A.:
\newblock {Modeling the Diffusion of New Durable Goods : Word-of-mouth Effect
  Versus Consumer Heterogeneity}.
\newblock In Laurent, G., Lilien, G., Pras, B., eds.: Research Traditions in
  Marketing.
\newblock Springer (1994)  201--229

\bibitem{Jaakola1996-CAA}
Jaakola, H.:
\newblock {Comparison and Analysis of Diffusion Models}.
\newblock In Kautz, K., Pries-Heje, J., eds.: {Diffusion and Adoption opf
  Information Technology}.
\newblock Chapman \& Hall (1996)  65--82

\bibitem{Geroski2000-MOT}
Geroski, P.:
\newblock {Models of Technology Diffusion}.
\newblock Research Policy \textbf{29}(4--5) (2000)  603--625

\bibitem{Bauckhage2013-MMO}
Bauckhage, C., Kersting, K., Hadiji, F.:
\newblock {Mathematical Models of Fads Explain the Temporal Dynamics of
  Internet Memes}.
\newblock In: Proc. ICWSM, AAAI (2013)

\bibitem{Bauckhage2014-CAT}
Bauckhage, C., Kersting, K., Rastegarpanah, B.:
\newblock {Collective Attention to Social Media Evolves According to Diffusion
  Models}.
\newblock In: Proc. WWW, ACM (2014)

\bibitem{Meyerson1957-NOA}
Meyerson, R., Katz, E.:
\newblock {Notes on a Natural History of Fads}.
\newblock American J. of Sociolog \textbf{62}(6) (1957)  594--601

\bibitem{Kullback1951-OIA}
Kullback, S., Leibler, R.:
\newblock {On Information and Sufficiency}.
\newblock Annals of Mathematical Statistics \textbf{22}(1) (1951)  79--86

\bibitem{Gradshteyn2007-TOI}
Gradshteyn, I., Ryzhik, I.:
\newblock {Tables of Integrals, Series, and Products}. 7th edn.
\newblock Academic Press (2007)

\end{thebibliography}

\end{document}